\documentclass[aps,prb,twocolumn,superscriptaddress,nofootinbib]{revtex4}

\usepackage{amsmath,amssymb}
\usepackage{graphicx}
\usepackage{epstopdf}
\usepackage{bm}
\usepackage{color}
\usepackage{mhchem}

\newcommand{\lsco}     {La$_{2-x}$Sr$_x$CuO$_4$}
\newcommand{\ybco}     {YBa$_2$Cu$_3$O$_y$}

\newcommand{\tso}      {$T_\text{SO}$}
\newcommand{\la}       {$^{139}$La}
\newcommand{\cu}       {$^{63}$Cu}
\newcommand{\laslr}    {$^{139}T_1^{-1}$}
\newcommand{\cuslr}    {$^{63}T_1^{-1}$}
\newcommand{\slr}      {$T_1^{-1}$}
\newcommand{\slrt}      {$(T_1T)^{-1}$}
\newcommand{\laslrt}    {$^{139}(T_1T)^{-1}$}
\newcommand{\cuslrt}    {$^{63}(T_1T)^{-1}$}
\newcommand{\htot}      {HTT $\rightarrow$ LTO}

\newcommand{\chiac}      {$\chi_\text{ac}$}

\begin{document}

\title{Low-energy spin dynamics and critical hole concentrations in \lsco\ 
	($0.07\leq x \leq 0.2$) revealed by $^{139}$La and $^{63}$Cu nuclear 
	magnetic resonance}  

\author{S.-H. Baek}
\email[]{sbaek.fu@gmail.com}
%\homepage[]{Your web page}
\affiliation{IFW-Dresden, Institute for Solid State Research,
Helmholtz Stra{\ss}e 20, 01069 Dresden, Germany}
\author{A. Erb}
\affiliation{Walther-Mei{\ss}ner-Institut, Bayerische Akademie der Wissenschaften,
Walther-Mei{\ss}ner-Stra{\ss}e 8, D-85748 Garching, Germany}
%\author{H.-J. Grafe}
%\affiliation{IFW-Dresden, Institute for Solid State Research, PF
%270116, 01171 Dresden, Germany}
\author{B. B\"uchner}
\affiliation{IFW-Dresden, Institute for Solid State Research,
Helmholtz Stra{\ss}e 20, 01069 Dresden, Germany}
\affiliation{Institut f\"ur Festk\"orperphysik, Technische Universit\"at 
Dresden, 01062 Dresden, Germany} 

\date{\today}

\begin{abstract}
	We report a comprehensive $^{139}$La and $^{63}$Cu nuclear magnetic resonance 
	study on \lsco\ ($0.07\leq x \leq 0.2$) single crystals.  
	The \la\ spin-lattice relaxation rate %\laslr\ 
	%reveals a sharp anomaly associated with the tetragonal to orthorhombic 
	%structural phase transition, whose position decreases linearly with increasing Sr  
	%doping $x$.  At low temperatures, 
	\laslr\ 
	is drastically influenced by Sr doping $x$ at low temperatures. 
	A detailed field dependence of \laslr\ at $x=1/8$ suggests that charge ordering 
	induces the critical slowing   
	down of spin fluctuations toward glassy spin order and competes with 
	superconductivity. On the other hand, the \cu\  
	relaxation rate \cuslr\ is well described by a 
	Curie-Weiss law at high temperatures, yielding the Curie-Weiss temperature 
	$\Theta$ as a function of doping. $\Theta$ changes sharply through a 
	critical hole concentration $x_c\sim 0.09$. $x_c$ appears to 
	correspond to the delocalization limit of doped holes, above which
	the bulk nature of superconductivity is established.  
%	Our relaxation and \textit{in situ} ac susceptibility data obtained at 
%	$x=0.2$ point the development of  
%	an inhomogeneous superconducting state in the vicinity of the 
%	structural phase boundary at zero temperature.     
%	Using our experimental observations, a revised phase diagram for 
%	\lsco\ is presented.  

\end{abstract}

%\pacs{76.60.Jx, 74.25.Ha, 76.60.-k, 74.72.Dn}

\maketitle

\section{Introduction}

\lsco\ (LSCO) is a prototype of the high-$T_c$ cuprates 
%and has been 
%intensively studied to date. Even with the most simple crystal structure, it 
that exhibits    
very rich structural, magnetic, and electronic phases whose 
understanding is believed to provide a key to understanding high-$T_c$ mechanism.  

Despite intensive studies on LSCO for three decades, however, there are still 
non trivial issues that are not fully understood.    
For example, the coexistence of magnetic and 
superconducting orders\cite{katano00, lake02, machtoub05, 
savici05, chang08} 
and an inhomogeneous superconducting state\cite{dordevic03} 
in the underdoped region have been known, 
but their underlying mechanism remains elusive.
The strongly enhanced magnetic  
order\cite{arai00,savici02,mitrovic08} observed near a hole concentration of $x=1/8$ is 
particularly interesting, as it appears to be linked to the stripe or the 
charge density wave (CDW) 
instability.\cite{takeya02,park11b,wu12}
%, which was verified  
%in other lanthanum cuprate family, \lbco\ and \lmsco\ 
%(M=Nd,Eu).\cite{tranquada95,fujita02,hucker07,fink09}
%
Recently, the CDW in \lsco\ was actually detected near $x=1/8$ by 
x-ray diffraction measurements,\cite{croft14,thampy14,christensen14}
and a high-field study of the Seebeck coefficient\cite{badoux16} 
reported that the CDW  
modulations cause the Fermi surface reconstruction in the limited doping range 
$0.085 < x <0.15 $, whose onset peaks at near $x=1/8$. It is noteworthy that 
the lowest doping limit $x=0.085$ for the CDW modulations agrees with the 
extrapolated value at  
zero temperature for the CDW onset probed by hard x-ray 
measurement.\cite{croft14}

With these latest experimental observations,  
several questions naturally arise: (1) how the CDW and 
superconductivity are related, (2) whether or how the CDW is
coupled to the enhanced spin order observed near $x=1/8$, and (3) why the CDW 
modulation appears at a considerably larger doping than the known critical 
doping $x\sim0.05$ for  
superconductivity,\cite{yamada98} above which 
drastic changes in the magnetic properties are 
observed.\cite{wakimoto99,fujita02a,li07a} 
%

%Another unsettled issue is the role of the structural instability for the rapid 
%suppression of bulk superconductivity in the overdoped region, near the 
%boundary between the low-temperature orthorhombic (LTO) and high-temperature 
%tetragonal (HTT) phases at $x\sim0.22$. While a direct relation between the 
%LTO structure and superconductivity is unknown, the fact that the phase 
%separation into SC and metallic regions near and above the LTO-HTT phase 
%boundary is inferred from experiments\cite{wen00,wakimoto05,wang07a,kaiser12} 
%is suggestive of the importance of 
%the crystal structure in the SC mechanism. 
% 

In this paper, we carried out \la\ and \cu\ spin-lattice relaxation rate (\slr)
measurements in superconducting \lsco\  
($0.07\leq x \leq 0.2$), in order to investigate how the normal and superconducting 
properties in \lsco\ evolve as a function of doped holes.   
%especially in light of the availability of the  
%high-quality single crystals with less disorder these days. 
%
%temperature $T_c$.
Based on the temperature and doping dependence of the \la\ relaxation rate, we 
propose that the unusual glassy behavior observed near $x=1/8$ may be  
considered as a fingerprint of charge order. The  
\cu\ relaxation measurements imply the 
presence of a critical hole  
concentration $x_c\sim0.09$, above which superconductivity emerges in a full 
volume fraction and the Curie-Weiss  
temperature $\Theta$ shares a similar doping dependence with $T_c$.   
%
%Our data further show that superconductivity becomes inhomogeneous in the 
%proximity of the structural instability.  

\begin{figure}
\centering
\includegraphics[width=0.9\linewidth]{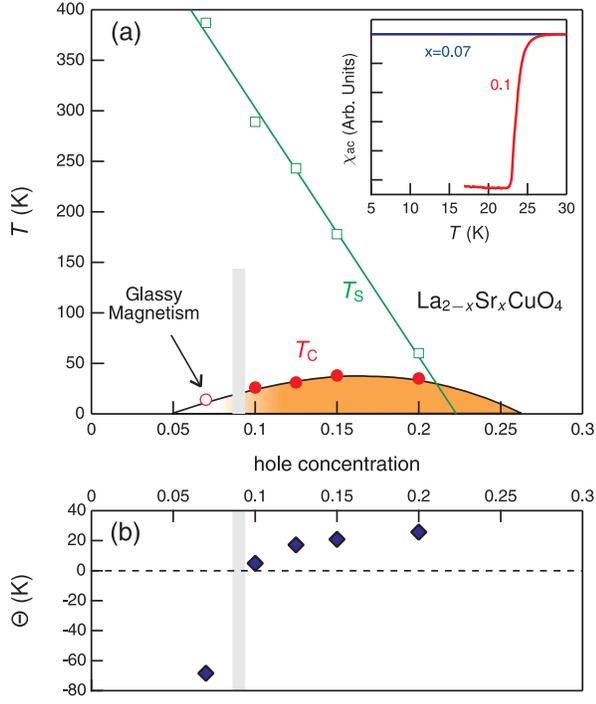}
	\caption{\label{pd} (a) $x$-$T$ phase diagram of \lsco. $T_S$ and $T_c$ 
	were all determined by our NMR measurements, except $T_c$ for $x=0.07$ 
	which was determined by a weak drop in dc susceptibility measurement. 
	As clearly shown in the inset, the SC transition for $x=0.07$ was not observed 
	in the \textit{in situ}
	ac susceptibility \chiac, in contrast to the sharp SC transition for $x=0.1$. 
	(b) The Curie-Weiss temperature $\Theta$ extracted from \cuslr\ measurements as   
	a function of doping $x$. Above $x_c = 0.09$, $\Theta$ abruptly changes and quickly 
	approaches a constant value. 
} 
\end{figure}

\section{Sample preparation and experimental details}

The  \lsco\ single crystals were grown with the traveling solvent floating zone 
method, as described in Ref. \onlinecite{baek12a}. 
We confirmed the 
superconducting (SC) transition temperature $T_c$ from the onset of 
the drop of the \textit{in situ} ac susceptibility \chiac\ in the nuclear 
magnetic resonance (NMR) tank circuit, 
and the resultant values are in good agreement with SQUID measurements.  
For $x=0.07$, a drop of \chiac\ was not detected
down to 4.2 K, in contrast to the sharp SC transition observed for $x\geq0.1$ 
(see the inset of Fig. \ref{pd}). 
This indicates that  
superconducting volume fraction is very small at $x=0.07$, but it becomes abruptly
100\% as Sr doping exceeds a critical value $x_c$ that is estimated to be 0.09 
(see Fig. \ref{pd}). 

\la\ ($I=7/2$) and \cu\ ($I=3/2$) NMR measurements were performed on  
\lsco\ single crystals with $x = 0.07$, 0.1, 0.125, 0.15, and 0.2 
in the range of temperature 4.2 --- 420 K and in an external field $H$ that ranges 
from 6 to 16 T. The crystallographic $c$ axis of the samples were aligned 
along the applied field direction using a goniometer.  
The spin-lattice relaxation rates \slr\ were measured
at the central transition ($+1/2 \leftrightarrow -1/2$) of both nuclei by monitoring 
the recovery of the echo signal after a saturating single $\pi/2$ pulse.  
Then the relaxation data were fitted to the appropriate fitting functions: For the
\cu,
\begin{equation}
\label{eq:cuT1}
	1-\frac{M(t)}{M(\infty)}=A\left(0.1e^{-(t/T_1)^\beta}+0.9e^{-(6t/T_1)^\beta}\right), 
\end{equation}
and for the \la,
\begin{equation}
\begin{split}
\label{eq:T1}
1-\frac{M(t)}{M(\infty)}=
&A\left(\frac{1}{84}e^{-(t/T_1)^\beta}+\frac{3}{44}e^{-(6t/T_1)^\beta}\right.   \\
+ &\left.\frac{75}{364}e^{-(15t/T_1)^\beta}+\frac{1225}{1716}e^{-(28t/T_1)^\beta}\right),
\end{split}
\end{equation}
where $M$ is the nuclear magnetization and $A$ is a fitting parameter that is
ideally one.  $\beta$ is the stretching exponent, which is less than unity when 
\slr\ becomes spatially distributed due to glassy spin freezing.  

\section{Experimental results and discussion}

\subsection{\la\ NMR}

\begin{figure}
\centering
\includegraphics[width=0.9\linewidth]{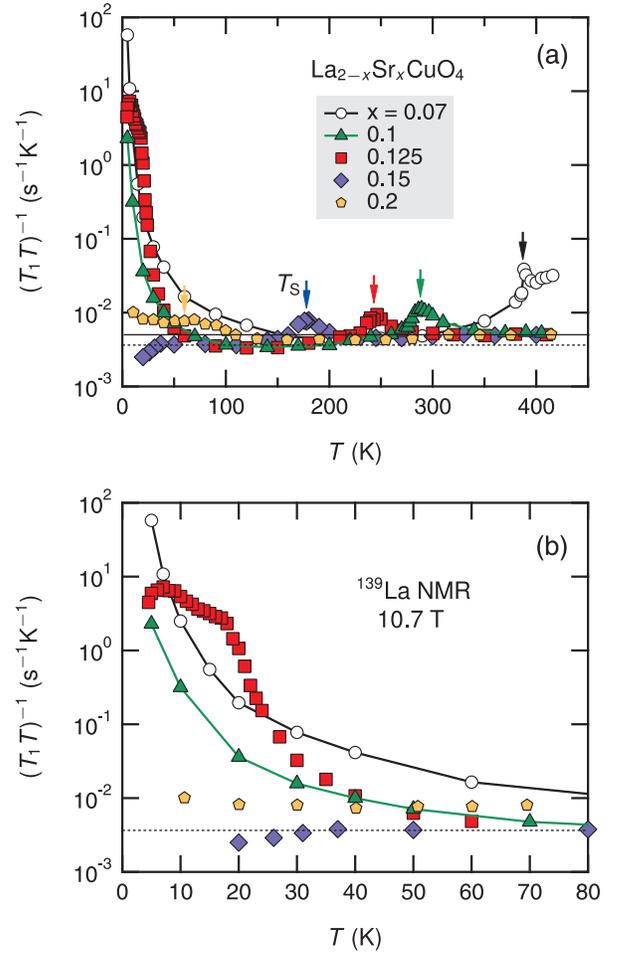}
	\caption{\label{la1} (a) Temperature dependence of
\laslrt\ as a function 
of doping $x$ measured at 10.7 T. The structural transition temperature $T_S$ 
	is identified from the sharp peak of \laslrt\ and denoted by the down arrows.  
The background 
\laslrt\ (solid horizontal line)
is nearly independent of $x$, which is slightly reduced
	below $T_S$ (dotted horizontal line). (b) At low temperatures, \laslrt\ 
	drastically changes depending on $x$. The 
	prominent enhancement of \laslrt\ for $x=1/8$ is clearly shown.} 
\end{figure}

Figure \ref{la1} shows the temperature and doping dependence of \laslrt\ measured 
at 10.7 T. The sharp anomaly of \laslrt\ is associated 
with the high-temperature tetragonal (HTT) to low-temperature orthorhombic 
(LTO) structural transition, giving rise to the \htot\ transition temperature 
$T_S$ which rapidly decreases with decreasing $x$. The 
background \slrt\ is nearly independent of $x$ [solid horizontal 
line in Fig. \ref{la1}(a)] in a wide temperature range except an enhancement 
observed at $T>350$ K for $x=0.07$ (see Ref. \onlinecite{baek12a} for  
discussion regarding a possible origin of the enhancement). The origin of 
the constant \laslrt\ is ascribed to the quadrupolar relaxation process due to 
the fluctuating electric field gradient (EFG).\cite{borsa89}
Interestingly,
after undergoing the structural transitions at $T_S$, the constant value of 
\laslrt\ drops slightly    
to another constant (dotted horizontal line). The 
small change of \laslrt\ subsequent to the \htot\ transition is accounted for by a 
local tilting of the EFG with respect to the $c$ axis,\cite{baek12a} that is, by the 
misalignment between the nuclear quantization axis and the applied field direction.

In the low temperature region, the temperature and doping dependence of 
\laslrt\ becomes extremely strong and complicated, as shown in Fig. \ref{la1}(b). 
For $x=0.07$, 
\laslrt\ is enhanced by more than three decades with decreasing temperature, representing 
the critical slowing down of SFs toward spin order.\cite{baek12a}
As $x$ is increased, this \laslrt\ enhancement is greatly 
suppressed by more than an order  
of magnitude at $x=0.1$ and disappears completely at optimal doping 
$x=0.15$. Strangely, \laslrt\ is enhanced below $T_c$ for slightly overdoped 
$x=0.2$, suggesting that an unusual spin dynamics arises at $x\sim 0.2$. We 
will return to this issue briefly in Section \ref{sec:cu}.

\begin{figure}
\centering
\includegraphics[width=0.9\linewidth]{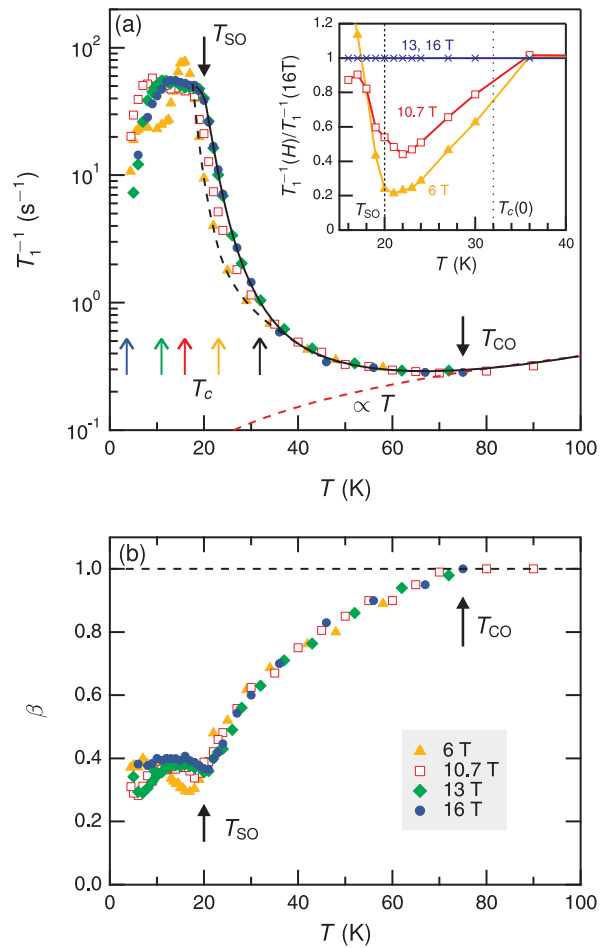}
\caption{\label{la2} 
(a) \laslr\ versus $T$ as a function of 
$H$ in LSCO:1/8 shows a strong field dependence. In particular, 
the \laslr\ upturn is suppressed with decreasing $H$, i.e., with 
	increasing $T_c$ 
	which is denoted by the up arrows. Inset: $T_1^{-1}(H)$ divided by 
	\slr\ at $H=16 T$. The significant suppression of \slr\ below $T_c(H=0)$ 
	is clearly shown.   
(b) Stretching 
exponent $\beta$ versus $T$ in LSCO:1/8. The  
deviation of $\beta$ from one occurs near 75 K. Clearly,
$\beta (T)$ is almost independent of $H$ except for the superconducting region.  
} 
\end{figure}

A predominant feature found in Fig. \ref{la1}(b) is the strong enhancement  
of \slrt\ with 1/8-doping, that is distinct from the data obtained at nearby 
dopings. Deviating at  
$\sim75$ K with respect to the temperature independent value, \slrt\   
rises sharply until it bends over at $\sim 20$ K.  Interestingly, it continues 
to increase before it drops abruptly at $\sim 8$ K. 
Note that the stretching exponent $\beta$ in Eq. (2) starts to  
deviate from unity below $\sim 75$ K, as shown in Fig. \ref{la2}(b). 
A $\beta$ value less than unity, which indicates a spatial distribution of 
\slr, is a key characteristic when \slr\ is strongly   
enhanced due to glassy spin freezing as observed $x=0.07$. 
%and, therefore, can be used as a measure for  magnetic inhomogeneity of a spin system.  
Therefore, from the temperature dependence of \slr\ as well as $\beta$, one 
can conclude that SFs are  
inhomogeneously slowed down below $\sim 75$ K at $x=1/8$, being connected with the 
enhanced glassy spin order detected in 
LSCO:0.12.\cite{savici02,savici05,hunt01,simovic03,mitrovic08}  

Such a spin-glass behavior at $x=1/8$ is surprising,
because it is caused by randomness that is usually absent in a metallic system. 
In lightly doped cuprates, the glassy 
behavior  
could be understood by randomly localized doped holes.\cite{cho92,chou93,niedermayer98}
However, near $x=1/8$, doped 
holes are largely delocalized and thus the origin of glassy spin  
order in LSCO:1/8 is hardly understood via the same mechanism as in 
the very underdoped regime.  
Furthermore, it was 
demonstrated that quenched disorder is not responsible for the  
glassy behavior in LSCO:0.12,\cite{mitrovic08} suggesting the relevance of 
1/8-anomaly for the glassiness.  
Therefore, we conjecture that 
charge stripe order, although it may be still rapidly fluctuating on the NMR time 
scale ($\sim\mu$s), may generate the randomness\cite{schmalian00,westfahl01} 
(e.g., localized holes) that could inhomogeneously slow down SFs.  

In order to check whether the inhomogeneous slowing down of SFs and charge order are 
related, we performed the field dependence of \laslr\ for LSCO:1/8, as shown 
in Fig. \ref{la2}. Clearly, the onset temperature at which \slr\ deviates  
from the $T$-linear behavior and $\beta$ deviates from unity is robust against 
the external field strength.  
Taken together with a recent NMR work in LBCO:1/8 
which strongly indicated that the slowing down of SFs occurs at the charge 
ordering temperature,\cite{baek15a} 
we believe that the anomalous change of both \slr\ and 
$\beta$ at $\sim75$ K is 
triggered by the charge ordering that is independent of 
$H$.\cite{hucker13} 
Remarkably, $T_\text{CO}\sim 75$ K is in excellent agreement with the  
static CDW ordering temperatures, 75 K (Ref. \onlinecite{croft14}) and 85 K 
(Ref. \onlinecite{christensen14}) for $x=0.12$ detected by x-ray 
diffraction studies.

Taking it for granted that charge ordering induces the inhomogeneous slowing 
down of SFs for 1/8-doped La cuprates, 
NMR might further probe the interplay between  
charge order and superconductivity. 
For that, we examine how the $T$-dependence of \slr\ is influenced by 
superconductivity [see Fig. \ref{la2}(a)]. 
At sufficiently high fields $\ge 13$ T, i.e. when superconductivity is   
nearly suppressed, it turns out that the high temperature side of the \slr\ 
peak is independent of $H$.
On the other hand, we find that the \slr\ upturn is clearly suppressed with 
decreasing field, i.e., with increasing $T_c$. 
We interpret that this behavior reflects a competition between  
charge order and superconductivity. 
If so, an immediate question is then why the reduction of the \slr\ 
upturn in Fig. \ref{la2} 
occurs well above the bulk $T_c (H)$.
The answer may be the presence of in-plane SC 
correlations above $T_c(H)$ which have been shown to  
persist up to $T_c(H=0)$ for moderate fields.\cite{li07, schafgans10,  
bilbro11, stegen13}
%This behavior may be due to 
%persistent in-plane SC correlations\cite{li07, schafgans10,  
%bilbro11, stegen13} above the bulk $T_c$ which can  
%coexist with charge order.\cite{berg09a}
This idea is substantiated by the fact that \slr\ is unaffected by external 
field above $T_c$ at zero field (the black up arrow in Fig. \ref{la2}).  

In comparison, as shown in Fig. \ref{la2}(b), $\beta$ is nearly  
independent of the magnetic field above \tso\ unlike \slr. This
indicates that although charge order itself competes with superconductivity, 
the spatial spin/charge inhomogeneity is robust  
against superconductivity.
When spins order below \tso, however,  
\slr\ as well as $\beta$ shows a significant magnetic field dependence.
While one may suspect that it arises from the interplay between superconductivity and spin 
order, it is unfortunate that the complex field dependence of \slr\ and $\beta$ 
does not allow a  
quantitative analysis. Regardless, our data reveals that there is a peculiar 
interplay of superconductivity and spin order.

\subsection{\cu\ NMR}
\label{sec:cu}

\begin{figure}
\centering
\includegraphics[width=0.9\linewidth]{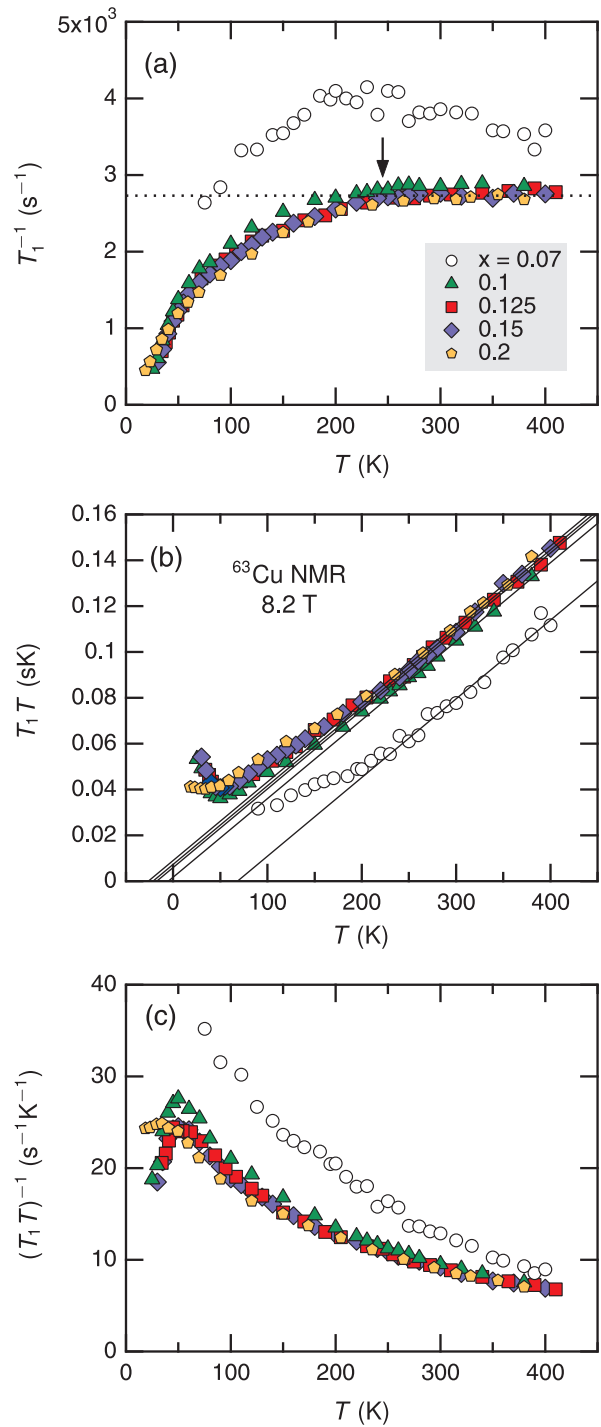}
\caption{\label{cuT1} (a) \cu\ \slr\ versus $T$ for different $x$ measured at 
	8.2 T applied along the $c$ axis.  \slr\ at $x=0.07$ is clearly distinct from  
	other data for $x\geq0.1$ which show almost $x$ independent behavior. (b) 
	Linear fits of $T_1T$ at high temperatures yield the Curie-Weiss temperature $\Theta$ 
	which abruptly changes near the critical doping $x_c\sim 0.09$. (c) 
	\cuslrt\ vs. $T$. For $0.1\leq x \leq 0.15$, \cuslrt\ forms a clear 
	peak at the same temperature $\sim 50$ 
	K. At $x=0.2$, the peak is notably suppressed.}
\end{figure}

The temperature and doping dependence of \cuslr\ 
measured at 8.2 T is shown in Fig. \ref{cuT1} (a).
For $x=0.07$,  
the relaxation is extremely fast, and the signal intensity rapidly 
decreases with decreasing $T$ due to the \textit{wipeout effect}.  
As $x$ 
is increased to a slightly larger doping 0.1, \cuslr\ is significantly reduced while 
roughly maintaining its overall temperature dependence. 
With further increasing doping, we find that \cuslr\ 
becomes nearly insensitive to $x$, especially at high temperatures where 
\cuslr\ is independent of temperature.\footnote{Our \cu\ NMR results are in 
marked contrast to the early  
results obtained in powder samples two 
decades ago\cite{ohsugi91,ohsugi94,itoh96} in which the $T$ dependence of 
\cuslr\ progressively changes with increasing $x$ up to the   
overdoped region. We believe that the previous NMR studies were  
strongly affected by substantial disorder in the samples, such as excess 
oxygen which is  
known to affect \laslr\ above 200 K in underdoped LSCO\cite{lascialfari93} 
and/or Sr impurity disorder.\cite{gooding97}}

If the spin-lattice relaxation is dominated by the staggered 
susceptibility $\chi(q=Q)$, where $Q$ is the AFM wave vector, it could be 
described by a Curie-Weiss law, i.e.,  
$(T_1T)^{-1}\propto \chi(Q) = C/(T+\Theta)$,\cite{ohsugi91,ohsugi94} where $C$ 
is the Curie constant  
and $\Theta$ the Curie-Weiss temperature.   
To confirm this behavior, we plot $T_1T$ versus $T$ as shown in Fig. \ref{cuT1}(b).
Indeed, at all dopings, a CW-like $T$ 
linear behavior was observed at high temperatures ($>250$ K). 
%Often, the temperature from which the CW behavior deviates was associated with the 
%pseudogap opening.\cite{suh00} However, whether the almost doping independent $T^*$ is 
%related to the pseudogap is far from clear.  
%In any case, 
The linear fits with the CW law in the 
high $T$ region give rise to $\Theta$. Here the Curie constant 
$C\propto J(J+1)$ was fixed, assuming the  
doping independent effective spin moment $J$.
The doping dependence of $\Theta$ is drawn in Fig. \ref{pd}. It shows 
that $\Theta$ parallels with $T_c$ as a function of doping in the region of  
$0.1\leq x \leq 0.2$, while 
it changes abruptly for $x=0.07$ where superconductivity was not 
detected by \chiac\ (see the inset of Fig. \ref{pd}). It should be noted that 
the strong wipeout effect at $x=0.07$ is  
similar to that observed in the underdoped region of \ybco\ (YBCO) below a critical hole 
concentration $p_c\sim 0.1$, in which the spin-glass 
phase coexists with superconductivity.\cite{baek12b}   This implies the 
existence of a critical    
hole concentration $x_c\sim 0.09$ below which glassy magnetism dominates 
over superconductivity, 
and above which AFM spin correlations increase sharply
and superconductivity fully replaces the magnetic volume fraction of the sample. 
Interestingly, the \la\ relaxation rate \laslr\ also shows an abrupt change at 
$T>300$ K as $x$ is increased through $x_c$ --- \laslr\ is strongly  
enhanced for $x=0.07$ with increasing temperature, being clearly distinguished 
from the data for $x\geq 0.1$ (see Fig. \ref{la1}).

The plot of \cuslrt\ vs. $T$ shown in Fig.~\ref{cuT1}(c) 
reveals another interesting feature.
\cuslrt\ increases with decreasing 
$T$, but drops below $\sim 50$ K ($> T_c$) forming a clear peak that is insensitive to $x$
in the range $0.1\leq x \leq 0.15$, which is qualitatively consistent with 
previous results.\cite{ohsugi91,ohsugi94,itoh96}  
The origin of the \cuslrt\ peak 
is unclear, but it may be either the rapid reduction of the AFM correlation length or 
a spin gap opening at low temperatures.\cite{itoh96}

The peak is, however, notably suppressed and moves to lower temperature at 
$x=0.2$. The abrupt suppression of the \cuslrt\ peak at $x= 0.2$ is actually 
consistent with a previous \cu\ NQR study in the powder sample.\cite{ohsugi94}  
Remarkably, at the similar doping $x\sim 0.2$, the in-plane 
resistivity $\rho_{ab}$ reveals a critical behavior.\cite{cooper09}  
$\rho_{ab}$ shows a $T$-linear  
behavior at all temperatures near $x=0.2$ just after the pseudogap vanishes at 
$x=0.19$, while it approaches purely $T^2$ 
behavior with either increasing or decreasing $x$.\cite{cooper09} 
Furthermore, as $x$ exceeds roughly 0.2, the system exhibits a Curie-like  
paramagnetism\cite{oda91,wakimoto05,kaiser12} and a 
drastic increase of the residual term $\gamma(0)$ of the specific heat.\cite{wang07a}
Therefore, we conclude that the suppression of the \cuslrt\ peak at $x=0.2$ 
reflects the intrinsic change of the physical properties of the system. A 
plausible explanation could be that the quasiparticle scattering mechanism critically changes 
beyond $x\sim 0.2$,\cite{cooper09} giving rise to an unusual spin dynamics at 
low energies. In fact, this could naturally account for the peculiar upturn of 
\laslrt\ in the SC state observed at $x=0.2$ (see Fig. \ref{la1}).

It is interesting to note that the $T$ dependence  
of \cuslrt\ remains nearly intact at low temperatures for $x\geq0.1$,  in 
stark contrast to the drastic  
change of that of \laslrt, in particular for $x=1/8$. It may be due to the fact that  
the relaxation of the \cu\ is three orders of magnitude faster than that 
of the \la\ at high temperatures [see Fig. \ref{la1}(a) and Fig. \ref{cuT1}(c)], and thus the 
\cu\ relaxation is not strongly affected by the  
slowing down of SFs associated with glassy spin freezing. In support of this, 
the stretching exponent $\beta$ for the \cu\ in Eq. (1) remains close to unity down to 
low temperatures for $x\geq 0.1$.

%Actually, this is corroborated by the parallel doping dependence of $\Theta$ and 
%$T_c$ (see Fig. \ref{pd}), suggesting that the AFM correlation and the 
%low-lying quasiparticle states associated with superconductivity may be interrelated.

\subsection{Phase diagram of \lsco}
\label{sec:phase}

Figure \ref{pd} shows a $T$-$x$ phase diagram determined from our NMR results. 
From the abrupt change of $\Theta$ (lower panel in Fig.~\ref{pd}), one could 
identify a critical hole concentration $x_c\sim 0.09$. Whereas the glassy 
magnetism dominates at $x<x_c$, causing the extremely fast upturn of \laslr\ 
(Fig. \ref{la1}) and the strong wipeout of the \cu\ signal at low 
temperatures, it is significantly suppressed and superconductivity  
soddenly appears in a full volume fraction as $x$ exceeds $x_c$.   
The underlying cause for the 
critical behavior at $x_c\sim0.09$ may be attributed to the localization limit of  
doped holes. For $x<  x_c$, doped holes are strongly localized and there are 
not enough carriers to be paired, accounting for the absence of the SC 
transition in \chiac\ for $x=0.07$. 
As $x$ reaches $x_c$, the metal-insulator transition takes place providing free charge 
carriers.\cite{muller98}

The critical hole concentration $x_c\sim 0.09$ is very close to the 
value of $\sim 0.1$ observed in YBCO,\cite{baek12b} suggesting that it could 
be the universal limit of delocalization of doped  
holes in cuprates. Furthermore, the fact that $x_c$ nearly coincides with the hole 
concentration $\sim 0.85$ above which the Fermi surface 
reconstruction is induced by the CDW modulations\cite{badoux16} may suggest that
the CDW and superconductivity are governed by the same criticality, despite 
the competing relationship between them inferred from the \la\ NMR.  

The phase diagram further shows 
that the structural transition temperature $T_S(x)$ goes to zero at $x\sim0.22$, 
which is in good agreement with previous  
studies.\cite{takagi92,yamada98} 
It is worthwhile to mention that, although the structural phase boundary exists 
inside the SC dome, it was demonstrated that superconductivity  
remains intact even if the structural boundary is shifted to much higher 
doping level by Pr-doping.\cite{schaefer94} Nevertheless, the peculiar 
behavior of the spin-lattice relaxation rates observed at $x=0.2$ suggests that 
the structural instability may have a large influence on the quasiparticle 
scattering process, resulting in an unusual low-energy spin dynamics.

\section{conclusion}

In summary, we presented \la\ and \cu\ NMR studies in a series of \lsco\ 
single crystals in the range $0.07\leq x \leq 0.2$. 
The unusual glassy 
behavior observed by the \la\ spin lattice relaxation \laslr\ at $x=1/8$ is 
ascribed to the randomness generated by charge  
order, which triggers the critical  
slowing down of spin fluctuations toward spin order. The field dependence of 
\laslr\ further suggests that the charge order competes with 
superconductivity.

The \cu\ spin lattice relaxation
\cuslr\ at high temperatures is governed by the staggered susceptibility
that obeys the Curie-Weiss law. Data show
that the Curie-Weiss temperature $\Theta$ changes abruptly at 
the critical doping $x_c\sim 0.09$ and becomes weakly dependent on further 
doping in the optimally doped regime, yielding a similar doping 
dependence of $\Theta$ and $T_c$.

\section*{Acknowledgement}
%We thank Markus H\"{u}cker for discussion.  
This work has been supported by the DFG Research Grant No. BA 4927/1-3.

\bibliography{mybib}

\end{document}